\documentclass[aps,prl,nofootinbib,10pt,tightenlines,notitlepage, twocolumn,floatfix,eqsecnum,superscriptaddress,showkeys]{revtex4-1}

\usepackage[utf8]{inputenc}          
\usepackage{graphicx}         
\usepackage[usenames,dvipsnames]{xcolor}
\usepackage{array,dcolumn,longtable} 
\usepackage{amsmath,amssymb,amsfonts,slashed} 
\usepackage{mathtools}          

\usepackage[linktocpage,breaklinks]{hyperref}

\usepackage{txfonts}
\usepackage{bm}
\usepackage{stmaryrd}
\usepackage{tensor}
\usepackage[utf8]{inputenc}

\usepackage{epsfig}
\usepackage{epstopdf}

\usepackage{cleveref}

\definecolor{mred}{RGB}{127,0,25}
\definecolor{mdgr}{RGB}{51,51,51}
\definecolor{mag}{RGB}{211, 54, 130}
\definecolor{verm}{RGB}{164, 25, 0}

\hypersetup{colorlinks=true,
            citecolor=NavyBlue,
            linkcolor=NavyBlue,
            urlcolor=NavyBlue}

\usepackage{siunitx}                                  
\DeclareSIUnit{\fm}{\femto\metre}                     

\begin{document}
\title{Neutron Star Equation of State in Light of GW190814}
\date{\today}

\author{Hung Tan}
\affiliation{Illinois Center for Advanced Studies of the Universe, Department of Physics, University of Illinois at Urbana-Champaign, Urbana, IL 61801, USA}
\author{Jacquelyn Noronha-Hostler}
\affiliation{Illinois Center for Advanced Studies of the Universe, Department of Physics, University of Illinois at Urbana-Champaign, Urbana, IL 61801, USA}
\author{Nico Yunes}
\affiliation{Illinois Center for Advanced Studies of the Universe, Department of Physics, University of Illinois at Urbana-Champaign, Urbana, IL 61801, USA}

\begin{abstract}
%
The observation of gravitational waves from an asymmetric binary opens the possibility for heavy neutron stars, but these pose challenges to models of the neutron star equation of state. We construct heavy neutron stars by introducing non-trivial structure in the speed of sound sourced by deconfined QCD matter, which cannot be well recovered by spectral representations. Their moment of inertia, Love number and quadrupole moment are very small, so a tenfold increase in sensitivity may be needed to test this possibility with gravitational waves, which is feasible with third generation detectors.

\end{abstract}
\maketitle

The LIGO/Virgo Collaboration (LVC) recently measured the coalescence of a black hole and a compact object with mass $M \approx (2.5-2.67) M_\odot$ \cite{Abbott:2020khf}.  Already much debate exists about whether the latter is a black hole, a primordial black hole, or a neutron star \cite{Most:2020bba,Broadhurst:2020cvm,Fishbach:2020ryj}. If the binary had had a mass ratio closer to unity, the event could have led to a measurement of the tidal deformability of the small compact object, perhaps providing direct evidence for whether it was a black hole or not. Lacking this measurement, the detection of the merger and post-merger phase, or an electromagnetic counterpart, arguments have been put forth that the small object has to be a black hole, since, after all, its mass is above what is currently believed to be the maximum mass of neutron stars $M_{\max} \approx (2.3,2.4) M_\odot$~\cite{Abbott:2020khf}. This belief is based on either galactic population modeling arguments~\cite{Alsing:2017bbc,Chatziioannou:2020msi}, which could suffer from selection bias, from mass threshold estimates with numerical relativity simulations~\cite{Margalit:2017dij,Rezzolla:2017aly,Ruiz:2017due,Shibata:2019ctb}, which make certain assumptions about the equation of state (EOS), or from LVC measurements of the EOS with the GW170817 event~\cite{Abbott:2018exr}, which use a particular spectral parameterization~\cite{Lindblom:2010bb,Lindblom:2012zi,Lindblom:2013kra}.

Inferences about the nature of a compact object that rely on prior assumptions on the EoS can be delicate given current nuclear physics uncertainties. Indeed, the community has had to revise its EOS assumptions several times before, as heavier and heavier neutron stars have been found, with perhaps the~$\sim2.1 M_\odot$ millisecond pulsars recently discovered being the latest example~\cite{Cromartie:2019kug,Buballa:2014jta}. The spectral EoS and other similar piecewise parameterizations do not directly model any nuclear microphysics, such as the possibility of deconfined Quantum Chromodynamics (QCD) matter within the core of the neutron star, which is expected at large enough baryon densities \cite{Alford:1997zt,Alford:1998mk,Alford:2001dt,Buballa:2003qv,Alford:2004pf,Fukushima:2003fw,Alford:2006vz,Alford:2002rj,Alford:2007xm,Dexheimer:2009hi,Dexheimer:2011pz,Alford:2013aca,Baym:2019iky,Annala:2019puf}. 
A key feature of all currently known models with deconfined QCD matter is that they present structure in their speed of sound, $c_s^2$, such as peaks, dips, kinks, and even discontinuities \cite{Bedaque:2014sqa,Alford:2015dpa,Ranea-Sandoval:2015ldr,Tews:2018kmu,Tews:2018iwm,McLerran:2018hbz,Jakobus:2020nxw,Annala:2019puf,Zhao:2020dvu}, which allow for a larger maximum mass. 
In fact, quarkyonic matter has been predicted to have a large peak in $c_s^2$ at sufficiently low baryon densities, which allows for neutron stars with large maximum masses~\cite{McLerran:2018hbz,Jeong:2019lhv}. In most cases, these features do not lead to mass twins, as the latter requires that the speed of sound remains zero for an extended region in the QCD phase space~\cite{Pang:2020ilf}. In fact, in our analysis we found it difficult to produce a mass twin that can reach a maximum mass as high as $M\geq 2.5 M_\odot$, but we leave further twin studies for a later paper.

\begin{figure*}[htb]
\centering
\includegraphics[clip=true,width=5.9cm]{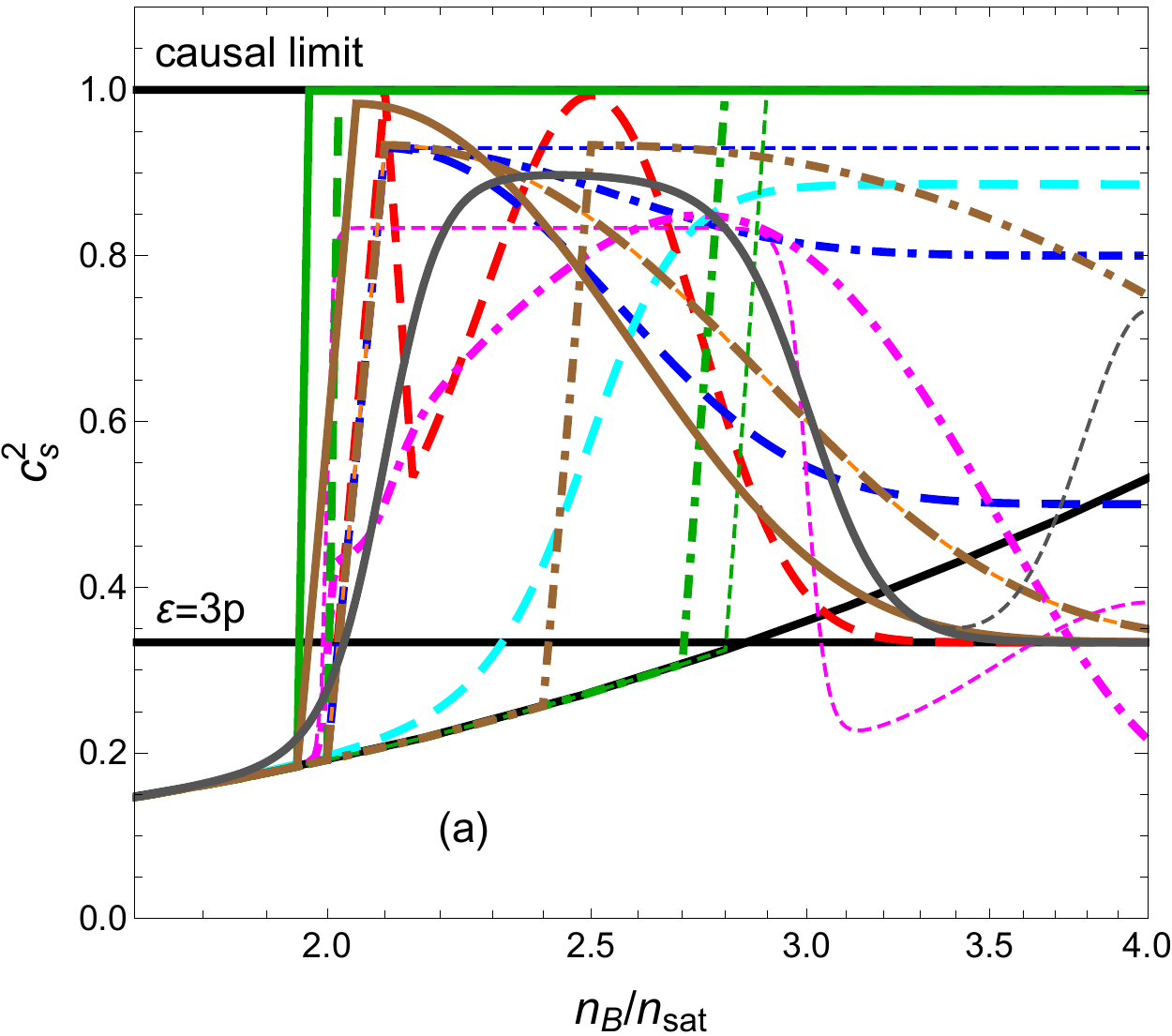} 
\includegraphics[clip=true,width=5.9cm]{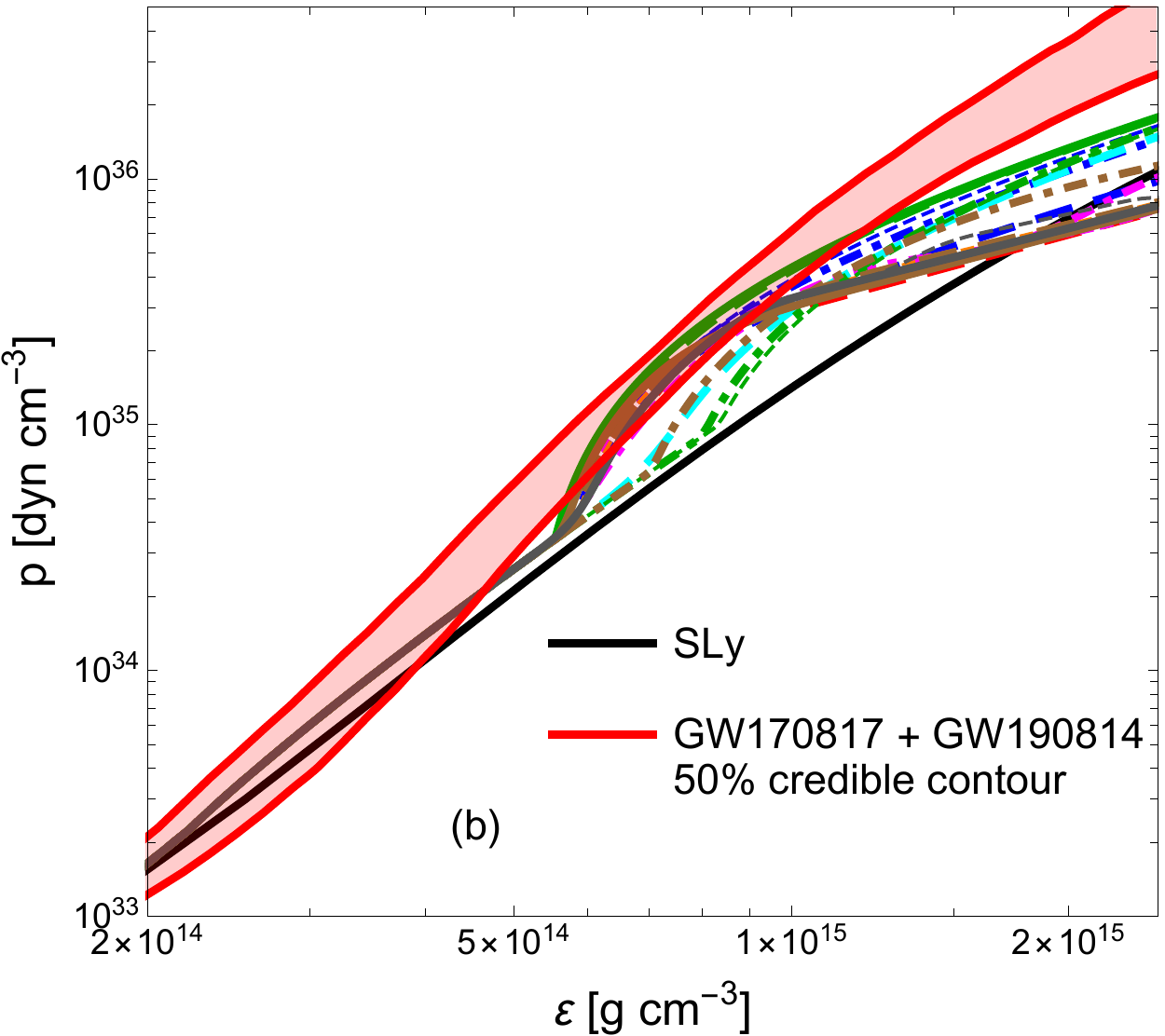} 
\includegraphics[clip=true,width=5.7cm]{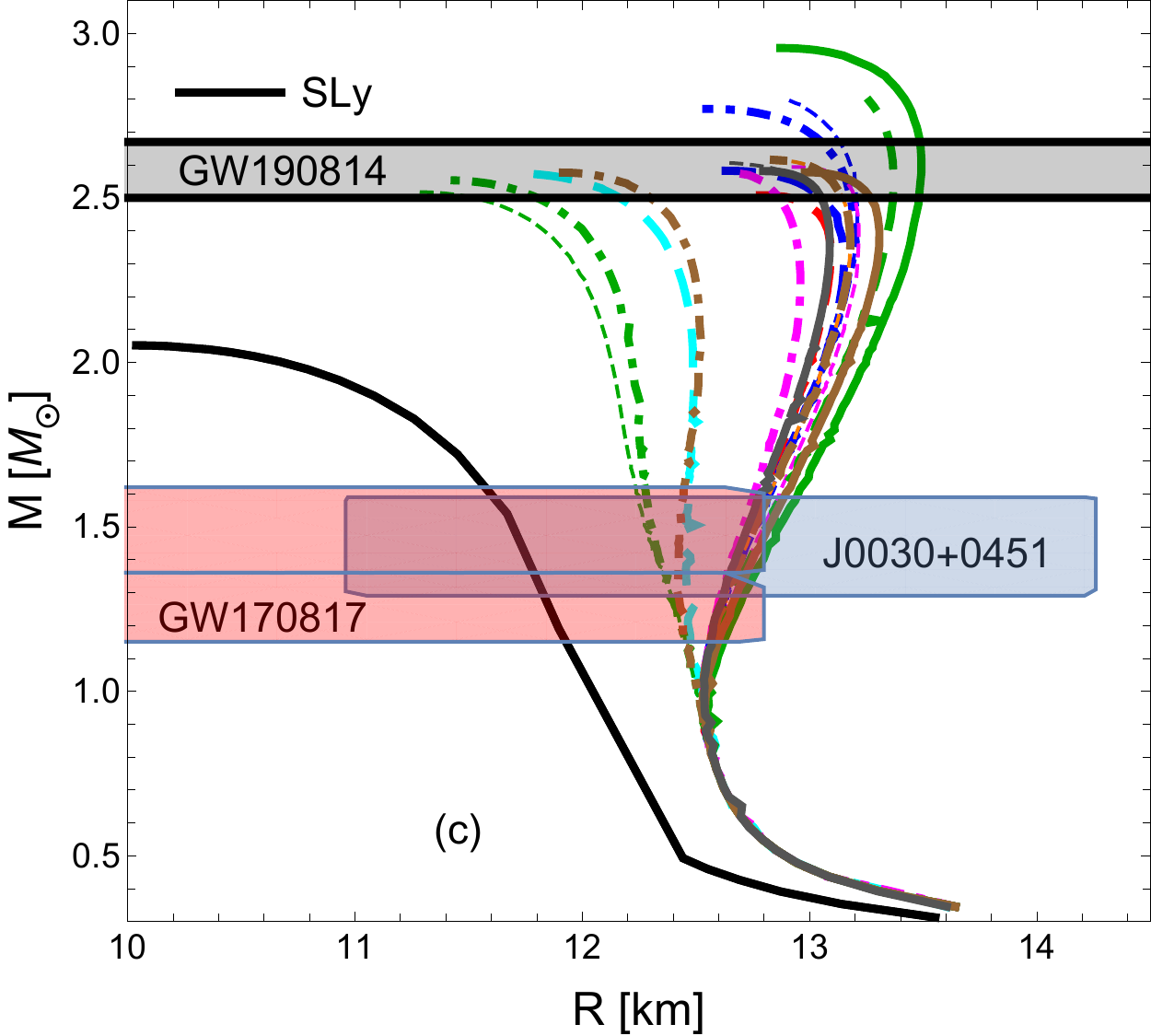} 
\caption{(Color online) 
 Parameterized speed of sound (a) as a function of baryon density. The individual colors indicate different choices for the functional form of $c_s^2$, as explained in the supplementary material. The horizontal lines denote the pQCD limit of $c_s^2=1/3$ and the causal limit $c_s^2=1$. Observe that all speed of sounds remain causal in the regime of interest. 
EOSs resulting from the parameterized speed of sounds (b). The shaded region corresponds to the 90\% confidence region reported by the LVC in~\cite{Abbott:2020khf}.
Mass-radius curves resulting from the new EOSs (c). Observe that all neutron star sequences reach a maximum mass of at least $M_{max}\geq 2.5M_\odot$. The recently inferred mass of the small compact object in GW190814 is shown in the gray band and the radius extracted from NICER observations of the isolated pulsar PSR J0030+0451 is shown in the blue square. 
 }\label{fig:cs2_summary}
\end{figure*}

\vspace{0.1cm}
\noindent \textit{Enlarging the Phase Space of EOSs}~--~A number of phenomenological methods exist to parameterize the EOS of neutron stars, with the three primary ones being piecewise polytropes~\cite{Read:2008iy,Annala:2017llu}, spectral functions~\cite{Lindblom:2010bb,Lindblom:2012zi,Lindblom:2013kra}, and parameterized $c_s^2$ functions~\cite{Alford:2015dpa,Ranea-Sandoval:2015ldr,Tews:2018kmu,Tews:2018iwm,Annala:2019puf}. Here we consider the latter with a wide variety of functional forms for $c_s^2$ that are able to capture the possible unique and kink-prone structure of the speed of sound inside neutron stars. For this first study, we leave the crust fixed, assumed to be given by the SLy EOS~ \cite{Chabanat:1997un,Douchin:2000kad,Douchin:2000kx,Douchin:2001sv} up to baryon densities of approximately $n\sim 2n_{sat}$, with $n_{sat}$ nuclear saturation density, and match it onto a chosen functional form for $c_s^2$ at larger densities. While some degree of error may exist from the crust assumption~\cite{Baym:2017whm}, this crust model is the same as that used by the LVC~\cite{Read:2008iy,Lindblom:2010bb}. Unlike in~\cite{Tews:2018kmu}, however, we purposefully do not ensure that derivatives of $c_s^2$ are continuous during the matching, precisely because we wish to model structure in the speed of sound. In fact, it is even possible to have jumps in the speed of sound if a phase transition occurs, or if new degrees of freedom become relevant at a specific baryon density. For instance, in~\cite{Parotto:2018pwx} the QCD critical point (at finite temperatures) and a first order phase transition are modeled by a 3D Ising model, which leads to a sharp peak followed by a dip in $c_s^2$ along the phase transition, while in~\cite{Jakobus:2020nxw} kinks are seen in $c_s^2$. 

Structure in the speed of sound, such as kinks, dips or peaks, can be understood in terms of a change in the degrees of freedom inside a neutron star, which in turn can be understood through the susceptibilities of the pressure, i.e.~the derivatives of the pressure with respect to the chemical potential $\mu_B$ ($\chi_n={d^nP}/{d\mu_B^n}$). For instance, when $n=1$, the susceptibility is simply the baryon density, $\chi_1=n_B = dP/d\mu_B$.   
The order of a phase transition is determined by the behavior of the $n^{th}$ susceptibility such that a first-order phase transition occurs when the baryon density jumps at some $\mu_B$, while a second-order phase transition occurs when $\chi_2$ diverges at some $\mu_B$. There is a direct connection between $c_s^2$ and $\chi_2$, given by $c_s^2={n_B}/({\mu_B \chi_2})$~\cite{McLerran:2018hbz}. Therefore, interesting structure in the speed of sound can provide direct insight into changes in the degrees of freedom within a neutron star. 

Structure in the speed of sound, however, cannot be added arbitrarily, since the resulting EOS must still respect certain restrictions. The restrictions we adopt are that the EoS (i) remain causal with $0<c_s^2\leq 1$, (ii) allow for $M_{\max}>2.5 M_\odot$, and (iii) fit within current radius and tidal deformability constraints from LVC and NICER measurements~\cite{Abbott:2018exr,Abbott:2018wiz,Riley:2019yda,Miller:2019cac}. While arguments from pQCD exist that $c_s^2$ should approach 1/3 from below at asymptotically large densities \cite{Kurkela:2009gj,Kurkela:2014vha,Hebeler:2013nza}, we do not restrict our functional forms to return to this limit. We make this choice because the pQCD limit may take place at extremely high densities that are well beyond the central density of neutron stars, which typically do not exceed $5$ times nuclear saturation density. 

An infinite family of EOSs satisfy the above restrictions, but we construct a representative set as follows. Below $1.5 n_{sat}$, the speed of sound is that of the SLy EoS, but at some chosen density $n_1 \geq 1.5 n_{sat}$, $c_s^2$ transitions (either through a linear or quadratic polynomial, or through a hyperbolic tangent) to a new regime, in which $c_s^2$ may have a bump or spike that eventually decays to a chosen value, or jumps to a large plateau, or oscillate about a constant value (see the Supplementary Material for more information). In some cases we combine multiple structures within the same EOS  with the idea that a neutron star may switch on many new degrees of freedom as large densities are explored. As we shall soon see, the common thread in all of these EOSs is a sharp rise between baryon densities of $n_B\sim (1.5-3) n_{sat}$, which produces a kink in the speed of sound and in the EOS, similar to what was seen in \cite{Annala:2019puf}, and which allows for very massive neutron stars.  Once our family of EOS are established, we solve the Tolman-Oppenheimer-Volkoff (TOV) equations to determine the mass-radius relation, and the Einstein equations in a slow-rotation expansion to deterime the moment of inertia, the Love number and the rotational quadrupole moment for a sequence of slowly-rotating and stable neutron stars in the Hartle-Thorne approximation~\cite{Hartle:1967he,Hartle:1968si}.


\vspace{0.1cm}
\noindent \textit{Kinky neutron stars}~--~Figure~\ref{fig:cs2_summary} shows the parameterized speed of sounds, the resulting EOSs and the mass and radius generated for each EoS for varying central densities. Strikingly, all of the EOSs that satisfy the aforementioned restrictions have a steep increase in $c_s^2$ at $n_1 \sim (1-3) n_{sat}$. The larger $n_1$, the wider the transition has to be to allow for stars with $M_{max}>2.5M_\odot$. A transition at $n_1 \gtrsim 3 n_{sat}$ does not produce an EOS stiff enough to allow $M_{max}>2.5M_\odot$.  Furthermore, if the transition occurs at $n_1 \sim (1-3) n_{sat}$, one can place a large multitude of structure at even larger densities, such as oscillations, without any obvious effect. This is because a large jump in the $c_s^2$ at low densities pushes the maximum central baryon density, associated with the maximum mass star, to lower values. Thus, we find that it is not possible to probe the EOS at $n_B/n_{sat} \gtrsim (4-5)$ because neutron stars in that regime would not be stable. The majority of our EOSs have a maximum central baryon density in the range of $n_B/n_{sat}\sim 2-4$ with a few extreme exceptions that can reach down to $n_B/n_{sat}\sim1.5$ or up to $n_B/n_{sat}\sim5$. Unsurprisingly then, our family of EOSs demonstrates a relationship between $n_1$ and the radius of the neutron star at its maximum mass: the star with a peak occurring at larger $n_1$ is more compact compared to the one with a peak at a smaller $n_1$ that is fluffier.

Figure~\ref{fig:cs2_summary} presents a subset of all the EOSs we investigated, with different colors (red, blue, cyan, magenta,  orange, dark green, brown, and dark gray). 
%
The green sub-family forces $c_s^2$ to rise sharply to the causal limit ($c_s^2 = 1$) at different values of $n_B/n_{sat}$. 
The cyan sub-family transitions to a constant value of $c_s^2$ through a hyperbolic tangent of varying steepness. 
The blue sub-family varies the end-point of $c_s^2$ at large densities , while the
orange sub-family varies the location of the peak. 
The brown sub-family varies both the peak location and width simultaneously in order to ensure that they all produce a maximum mass of $M\geq 2.5 M_\odot$. 
The dark gray sub-family has the same initial peak structure, but varies the functional form of $c_s^2$ after this peak. 
The magenta sub-family includes oscillations after an initial rise, and the red sub-family includes a double peak structure. 

\begin{figure}[htb]
\centering
\includegraphics[clip=true,width=8cm]{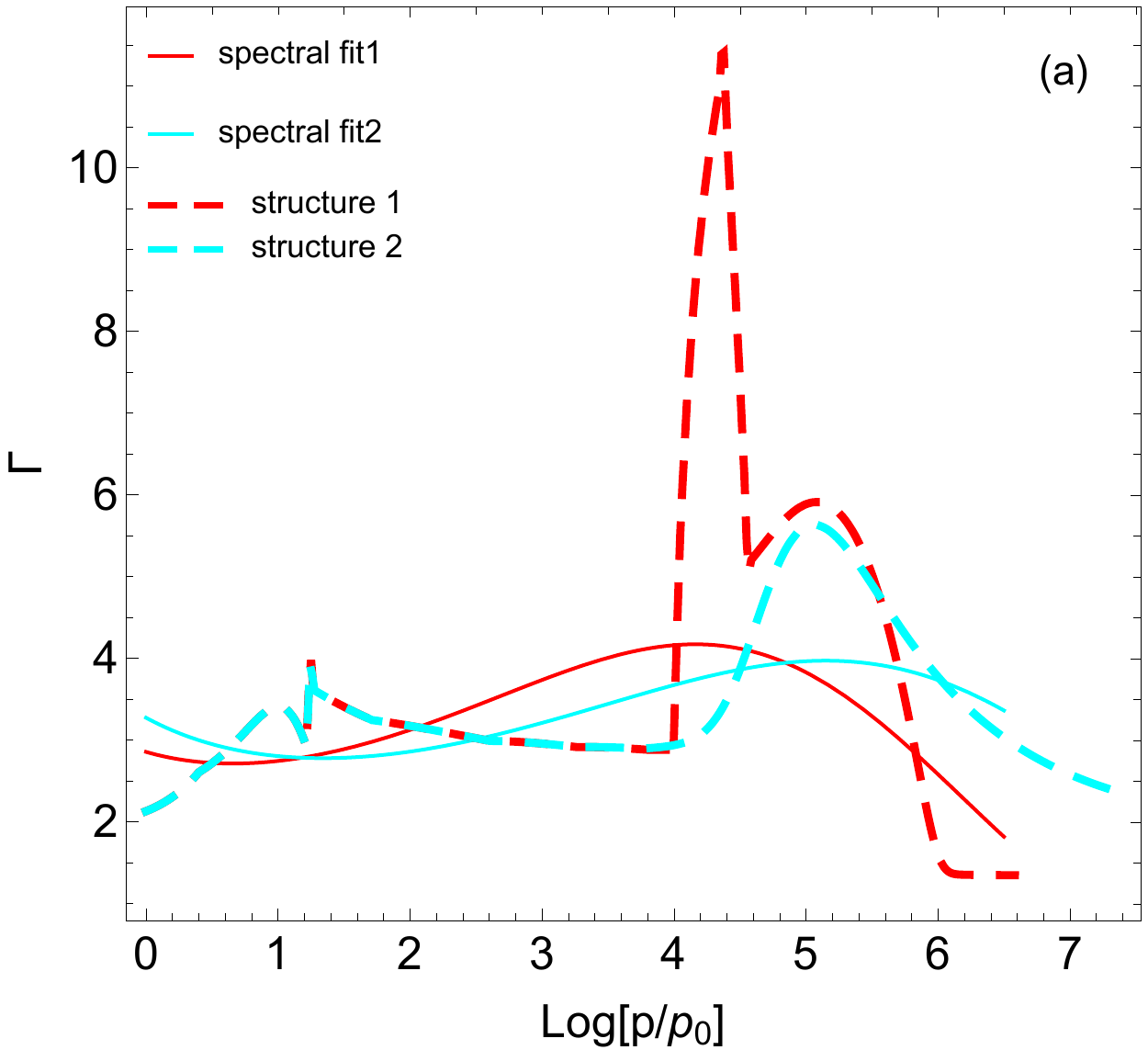} 
\includegraphics[clip=true,width=8cm]{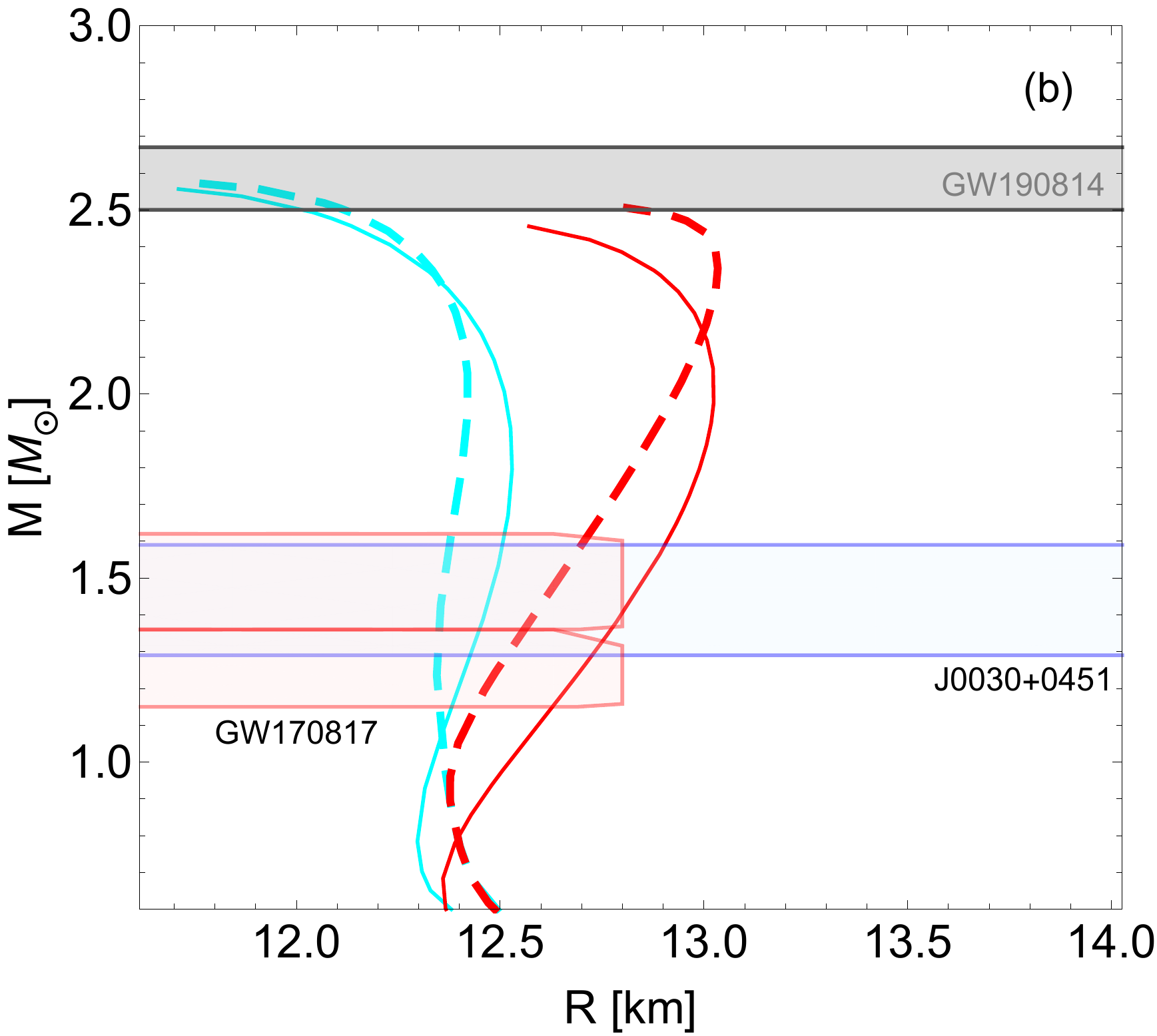} 
\caption{(Color online) (a) $\Gamma(p) := [(\epsilon + p)/p] dp/d\epsilon$ as a function of  $\log(p/p_0)$, with $p_0$ the pressure when the baryon density equals half nuclear saturation, $p$ the pressure and $\epsilon$ the energy density, and (b) mass-radius curves. In both panels, we use 2 members of the EOS family we showed in Fig.~\ref{fig:cs2_summary} (dashed red and cyan lines, following the same color and line styles as in that figure), as well as spectral EOS fits to these members using the LIGO prior range (dotted lines). The spectral fit is able to reproduce $\Gamma$ on average, but it misses the sharp features in the speed of sound. These missed features lead to large deviations in observables, such as in the mass-radius curves.  
}\label{fig:Gamma}
\end{figure}

Although the spectral EOS parameterization of~\cite{Lindblom:2010bb,Lindblom:2012zi,Lindblom:2013kra} can fit a wide range of EOSs, including the family introduced here to better than $5\%$, the spectral parameterization is not capable of fitting sharp structure in the speed of sound. We can see this in Fig.~\ref{fig:Gamma}, which shows $\Gamma(p) := [(\epsilon + p)/p] dp/d\epsilon$, where $\epsilon$ is the energy density and $p$ is the pressure, and the mass-radius relation, both computed with the spectral EOS parametrization (with a crust attached at half nuclear saturation density) fitted to two members of the EOS family of Fig.~\ref{fig:cs2_summary}. The spectral fit goes through an average of all the structure in the EOS, a feature that was known already from the original work that introduced this parameterization~\cite{Lindblom:2010bb}. In our case, however, this is problematic because it is this precise structure that allows for neutron stars with masses above $2.5 M_\odot$. 
As we can see on the low panel of this same figure, the error incurred by the spectral representation translates into a different shape of the mass-radius relation, which no longer fits within the maximum mass constraint.
%

One may wonder why the EOS family we study here lies outside the 50\% credible region found by the LVC (as shown in Fig.~\ref{fig:cs2_summary}), when the spectral representation is able to produce a good fit to our EOS family. The answer is that the spectral representation produces a good fit \textit{on average}, but is not capable of reproducing fine structure in the speed of sound. As usual in Bayesian analysis, the 50\% credible region in Fig.~\ref{fig:cs2_summary}) should be understood as the credible region for a spectral EOS prior, and not in the context of the true EOS of nature, whatever that may be. Since the spectral EOS (with 4 parameters) cannot reproduce kinks, deeps or peaks in the EOS, its 50\% credible region will be different than that obtained if one were to re-analyze the data with an EOS that was able to reproduce such structure. 

\begin{figure*}[htb]
\centering
\includegraphics[clip=true,width=5.9cm]{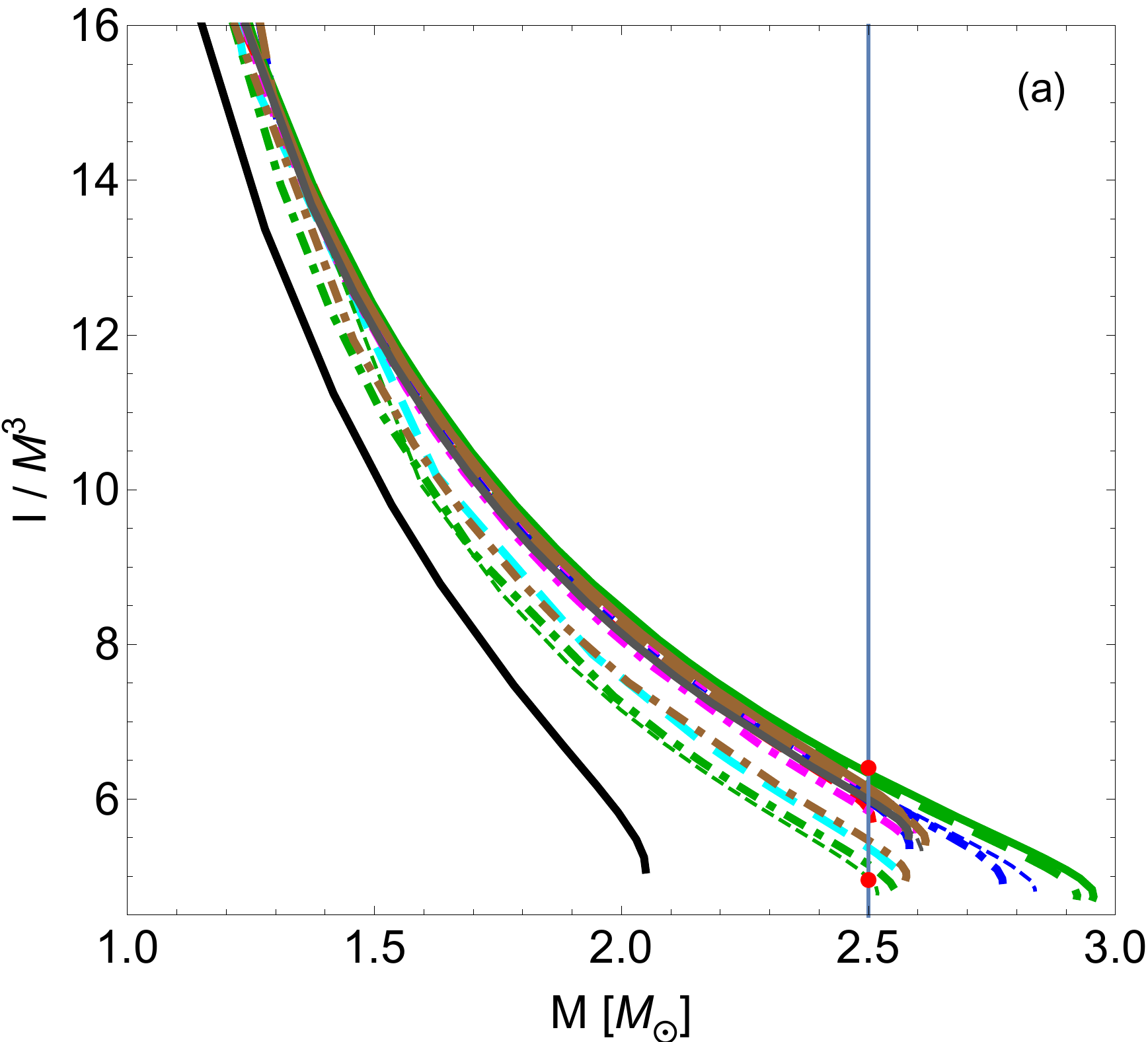}
\includegraphics[clip=true,width=5.9cm]{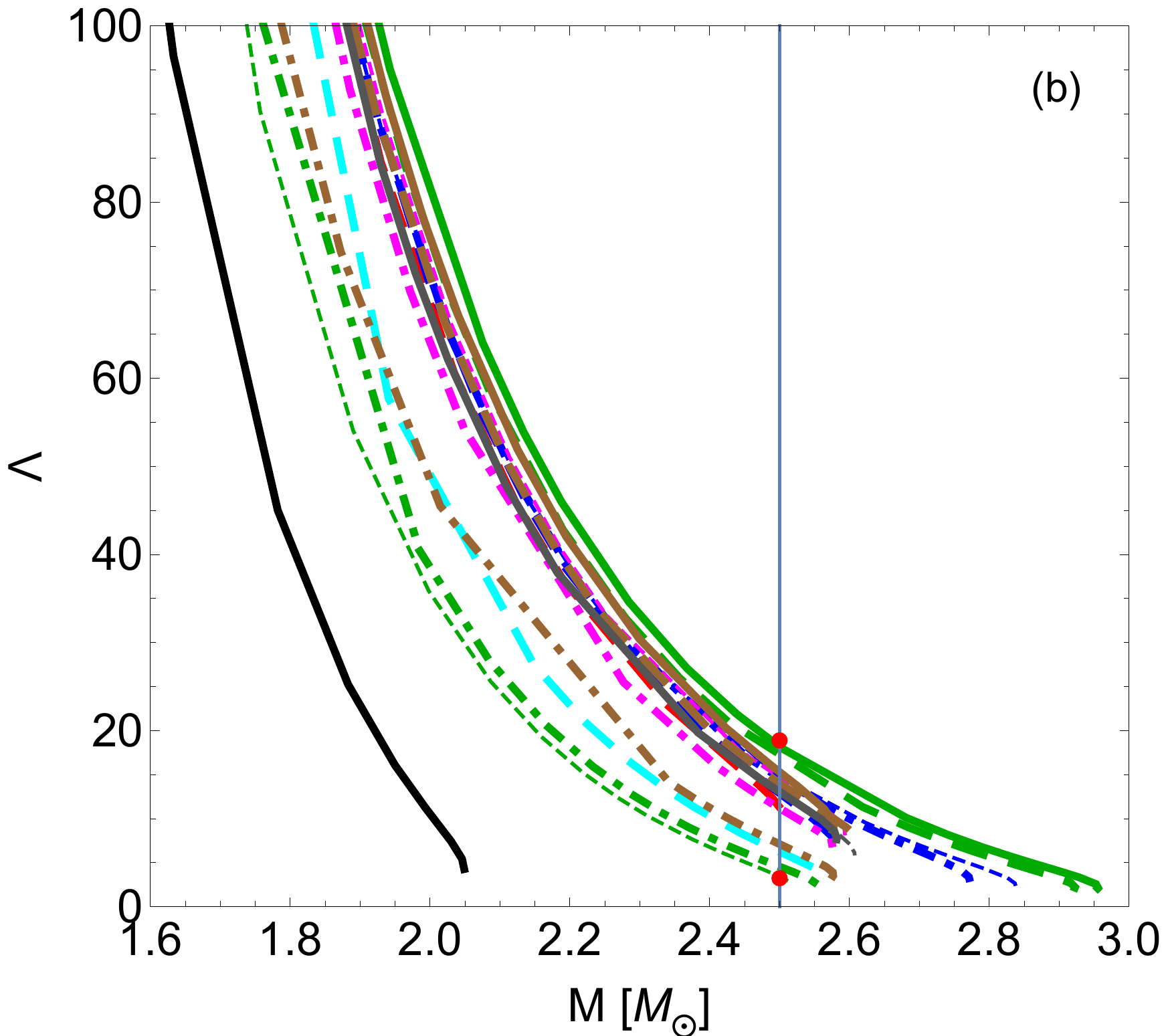}
\includegraphics[clip=true,width=5.9cm]{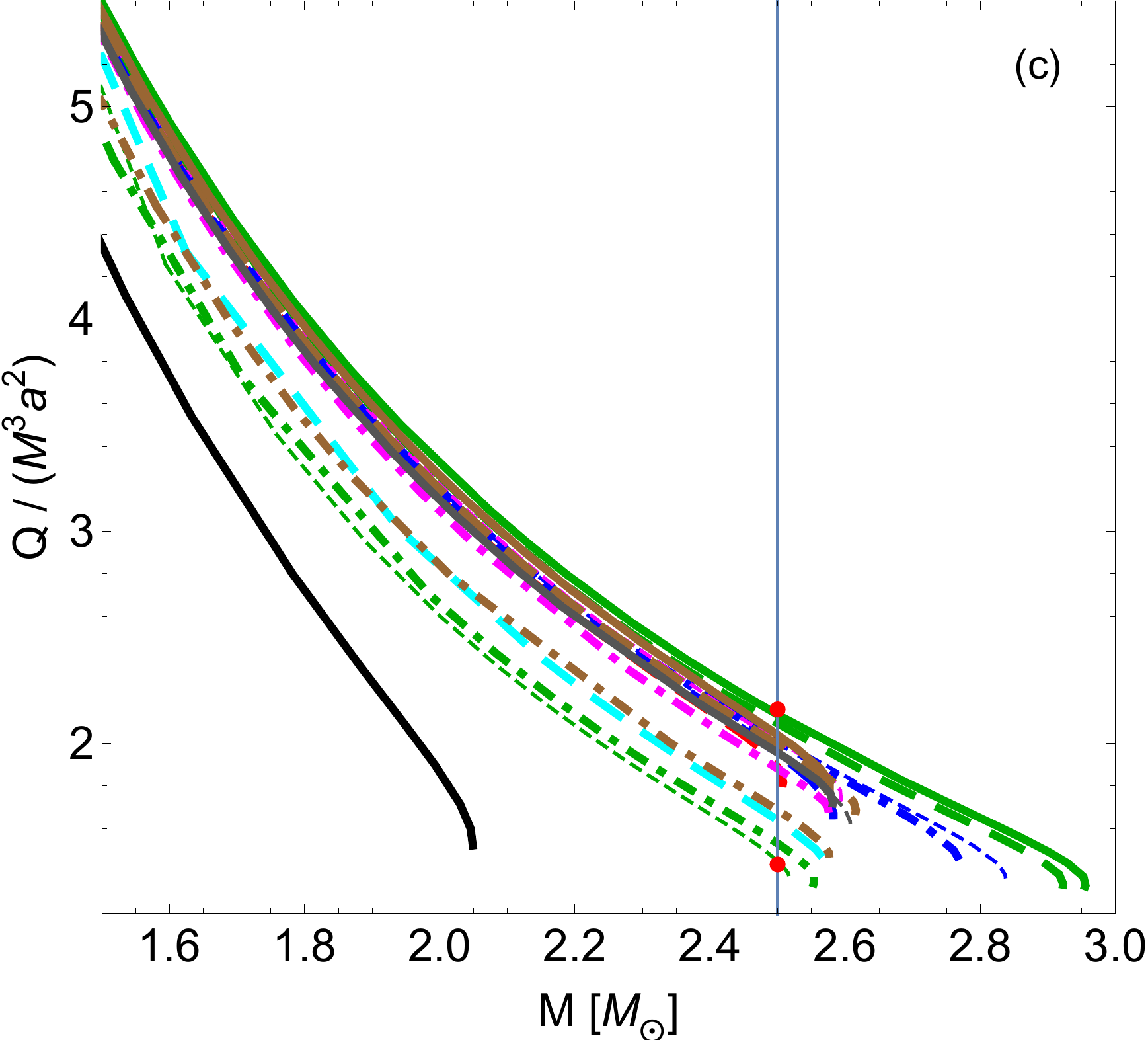}
\caption{(Color online) Dimensionless moment of inertia $I/M^3$ (a), tidal deformability parameter $\Lambda$ (b), and rotational quadrupole moment $Q/(M^3 a^2)$ (c) versus mass for our EOS family. The red dots indicated the values of $I/M^3$, $\Lambda$ and $Q/(M^2 a^2)$ for a $2.5 M_\odot$ star. Observe that the tidal deformability decreases with mass, and in particular, it drops to $\Lambda < 35$ for $M>2.5 M_\odot$.
}
\label{fig:GW_dim}
\end{figure*}

Given the family of EOSs we constructed here, we can also compute other neutron star observables that may be within reach in the near future, such as the moment of inertia $I$, the rotational quadrupole moment $Q$ and the ($\ell = 2$, electric-type) tidal Love number $\lambda_2$. Figure~\ref{fig:GW_dim} presents $I$, $Q$, and the tidal deformability parameter that enters the LVC waveform model $\Lambda$, which is linearly related to $\lambda_2$, for our EOS family. All quantities are non-dimesionalized through the mass $M$ and the dimensionless spin parameter $a := S/M^2$, with $S$ the (magnitude of the) spin angular momentum. As expected, all of these quantities decrease with $M$, reaching very small values when $M>2.5 M_\odot$. For example, when $M=2.5 M_\odot$ (red dots in Fig.~\ref{fig:GW_dim}), the moment of inertia reaches values of $I/M^3 \in (4.95,6.4)$, while the rotational quadrupole moment reaches values of $Q/(M^3 a^2) \in (1.43,2.16)$, and the tidal deformability reaches values of $\Lambda \in (3.23,18.88)$. The low end of these ranges are very close to what $I$, $\Lambda$ and $Q$ would be for black holes, implying these non-rotating stars are as close as they could be to the black hole limit, while remaining stable.

The small values of the above observables for very heavy neutron stars has important implications on their detectability through gravitational wave inspiral observations. For gravitational wave observations of the late inspiral phase to be able to measure such small values of $\Lambda$ or $\kappa = -Q/(M^3 a^2)$  they must have a resolution of $|\delta \Lambda| \lesssim 10 \gtrsim |\delta \kappa|$. Current gravitational wave measurements, using for example the GW170817 event~\cite{TheLIGOScientific:2017qsa}, can resolve $\Lambda$ for a $1.4 M_\odot$ star to roughly $\delta \Lambda \approx (100-400)$ at 90\% confidence, while $\kappa$ is correlated with a certain combination of the object's spin, yielding no measurements to date. This implies that one needs at least a one order of magnitude increase in the signal to noise ratio to be sensitive to the small tidal deformabilities of very heavy neutron stars. This could be accomplished with third-generation detectors, such as Cosmic Explorer or Einstein Telescope~\cite{Carson:2019rjx,Ap_Voyager_CE,ET}, or with a fortuitous close event when aLIGO reaches design sensitivity.

The observables presented above show a relative fractional variability\footnote{Fractional variability is defined as $\delta := [1 - X(M)/(\left<X\right>(M))]$ for any function $X(M) $ with $\left<X\right>(M)$ its average with respect to all EOSs considered.} of about $40\%$ with respect to changes in the structure of the EOS, but this variability can be essentially eliminated by constructing the I-Love-Q relations~\cite{Yagi:2013bca,Yagi:ILQ}. When doing so, we find relative fractional variabilities of $\lesssim 1\%$. This result is important because it implies that another set of universal relations must also exist, namely that between the tidal deformabilities $\lambda_2^{(1)}$ and $\lambda_2^{(2)}$ of two heavy neutron stars in a compact binary~\cite{Yagi:2015pkc}. If these binary Love relations exist for heavy neutron stars, one could then use the EOS-independent method employed on GW170817 to infer the mass and radius of such stars~\cite{Abbott:2018exr}, which does not require a prior choice of EOS parameterization. 

\vspace{0.1cm}
\noindent \textit{Future Directions}~--~We have demonstrated that, if GW190814 was generated by the coalescence of a black hole and a neutron star, then the neutron star EOS is highly likely to contain non-trivial structure in $c_s^2$ between $n_B/n_0\sim 1.5-3$ to almost the causal limit. This result is important because it would imply a large change in the degrees of freedom at zero temperatures and low baryon densities in the QCD phase diagram - possibly even the presence of a phase transition. Proving that this is the case, however, requires the measurement of very small tidal deformabilities in the inspiral phase, which would only be accessible with third-generation gravitational wave detectors, or alternatively the detection of the merger and post-merger phase~\cite{Most:2018eaw}.  

While in this study we apply an agnostic approach to the degrees of freedom within the core of a neutron star, it is likely that only models that have drastic changes in the degrees of freedom (such as deconfined QCD matter) can produce $M_{max}\geq 2.5 M\odot$ while preserving causality.  For instance, the quarkyonic phase \cite{McLerran:2007qj} naturally leads to this behavior \cite{McLerran:2018hbz,Zhao:2020dvu}, which would have far reaching consequences not only for cold neutron stars but also for the finite temperature QCD phase diagram probed at the Beam Energy scan at RHIC, and during the the merger of neutron stars themselves \cite{Most:2018eaw}.  

Future work could consider whether there are any other ways to produce such massive neutron stars, for example through a very stiff crust model that yields an EOS that remains causal. With that analysis in hand, one could then construct an improved spectral parameterization that is capable of capturing kinks in the EOS, because detecting these would provide invaluable information about the state of matter at densities above nuclear saturation. Other work could focus on extending the methodology studied here to produce mass twins \cite{Alford:2013aca,Dexheimer:2014pea,Benic:2014jia,Montana:2018bkb,Jakobus:2020nxw,Pang:2020ilf}, and to study whether other nuclear physics properties can be extracted from gravitational wave observations~\cite{Fattoyev:2017jql,Carson:2018xri,Capano:2019eae,Zimmerman:2020eho,Carson:2019xxz}. Once the merger and post-merger phases of the coalescence of neutron stars becomes detectable, hopefully when advanced LIGO reaches design sensitivity, one may also search for signatures of heavy neutron stars through their tidal disruption, mass ejecta, and kilonova features.  

{\emph{Note added after submission:}} A paper~\cite{Fattoyev:2020cws} has recently argued that heavy-ion data would exclude the potential equations of states we have constructed. This argument is flawed because the results it is based on, those of~\cite{Danielewicz:2002pu}, are over 20 years old today, and thus, superseded by new heavy-ion results.
Reference~\cite{Danielewicz:2002pu} uses an old model that allows only for hadronic degrees of freedom to analyze low-energy, heavy-ion data from the early 2000s. The model does not employ first-principle Lattice QCD calculations because the data is at beam energies that are well outside the regime of validity of the former~\cite{Bazavov:2017dus,Borsanyi:2018grb,Noronha-Hostler:2019ayj,Monnai:2019hkn}. Instead, the model used in~\cite{Danielewicz:2002pu} employs a hardonic-only approximation, without allowing for quark and gluon degrees of freedom. A recent analysis of recent HADES data~\cite{Adamczewski-Musch:2019byl} finds average temperatures of $T>70$ MeV, which may well be within a deconfined state of matter \cite{Critelli:2017oub}, and in fact, another initial study has recently found that deconfined matter is preferred by the data at these beam energies~\cite{Spieles:2020zaa}. Moreover, even the hadronic-only approximation used in~\cite{Danielewicz:2002pu} is at odds with recent hadronic-only calculations with an updated transport code~\cite{Hillmann:2019wlt}. A state of the art model would require an event-by-event, relativistic, viscous, hydrodynamic code with BSQ conserved charges, because viscous effects are  non-negligible and alter the path through the phase diagram \cite{Dore:2020jye}.
On top of this, heavy-ion collisions reach high temperatures, and thus, inferences on the equation of state at these energies should be compared to those obtained from observations of the merger of neutron stars, which can also reach $100$ MeV~\cite{Adamczewski-Musch:2019byl}. During the inspiral, neutron stars are at much lower temperatures and, thus, the equation of state inferred from the tidal deformability during the inspiral is in a different regime of the QCD phase space.

\vspace{0.1cm}
\noindent \textit{Acknowledgments}~--~The authors would like to thank Veronica Dexheimer, Hank Lamm, and Mauricio Hippert for useful discussions related to this work.  J.N.H. acknowledges the support of the Alfred P. Sloan Foundation, and support from the US-DOE Nuclear Science Grant No. DE-SC0019175. H.~T.~and N.~Y.~acknowledge support from NASA Grants No. NNX16AB98G, 80NSSC17M0041 and 80NSSC18K1352 and NSF Award No. 1759615. The authors also acknowledge support from the Illinois Campus Cluster, a computing resource that is operated by the Illinois Campus Cluster Program (ICCP) in conjunction with the National Center for Supercomputing Applications (NCSA), and which is supported by funds from the University of Illinois at Urbana-Champaign.

\bibliography{BIG}


\newpage
\clearpage
\section{Supplemental Material}

\begin{figure}[htb]
\centering
\includegraphics[width=1\linewidth]{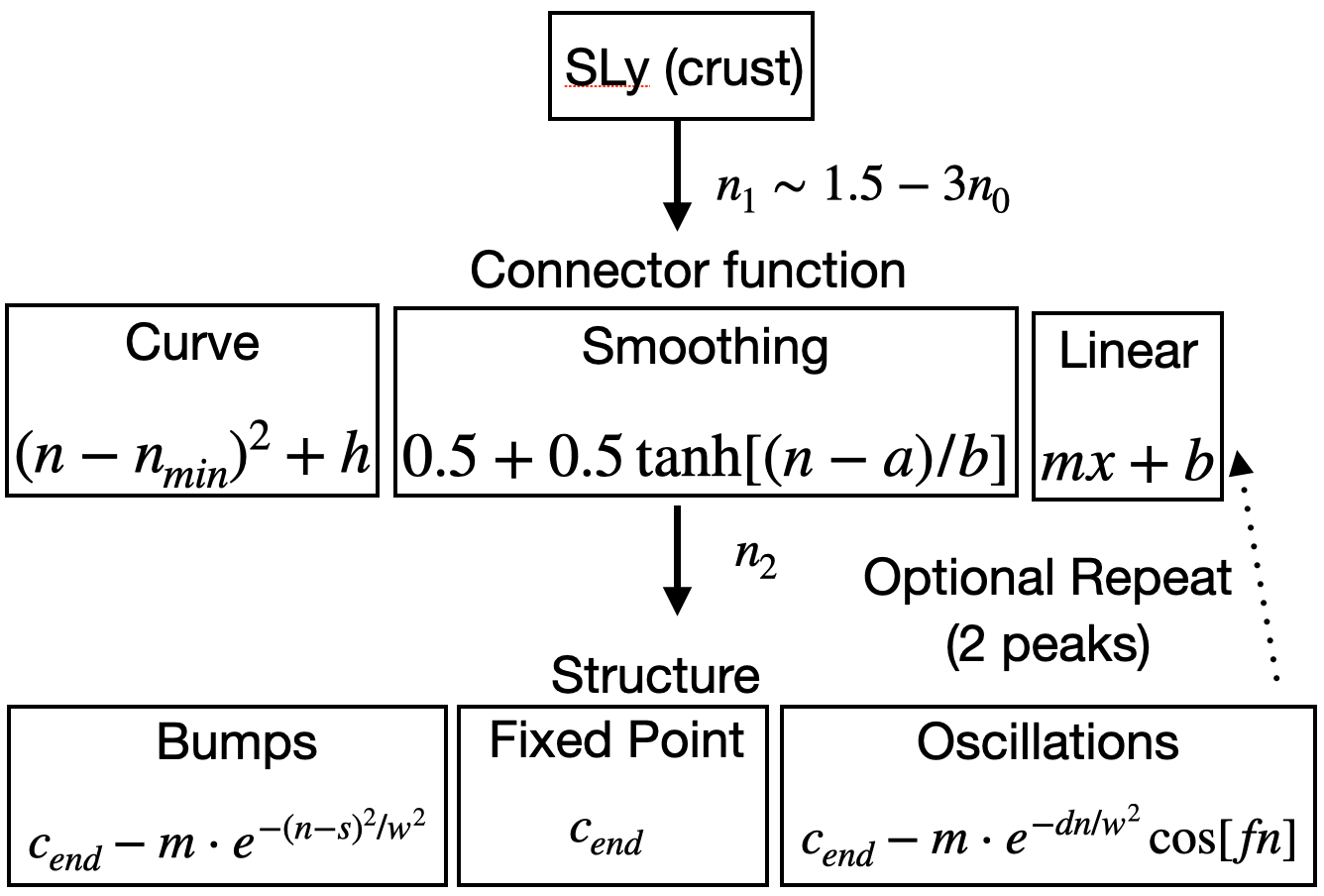} 
\caption{Flow chart of the piecewise functional forms of $c_s^2$ to create an EOS. More complicated structure in $c_s^2$ can be created with multiple connector and structure functions.}\label{fig:flow}
\end{figure}

The method to construct the family of EOSs we study here is illustrated in Fig.\ \ref{fig:flow}. We divide the EOS into 2 regimes, a low density regime $n_B < n_1$ and a high density regime $n_B > n_2 > n_1$, connected through an intermediate regime in $n_1<n_B<n_2$. In the low density regime ($n_B<n_1$), we use the SLy EOS \cite{Chabanat:1997un,Douchin:2000kad,Douchin:2000kx,Douchin:2001sv} to represent nuclear matter below nuclear saturation density. In the high density regime ($n_B>n_2>n_1$), we use a structure function that can introduce oscillations or bumps in the speed of sound. For instance, the structure function can introduce a bump or a gradual downward trend through a Gaussian function, it can introduce a fixed point where the speed of sound has a constant value, or it can introduce oscillations through trigonometric functions. In the intermediate regime ($n_1<n_B<n_2$), we use a connector function that can provide interesting structure (kinks, dips, etc) in the speed of sound.  For instance, we can generate a spike in the speed of sound by combining two linear connector functions, or by smoothly connecting a large jump through a hyperbolic tangent function. Below we detail one specific subset of EOS to provide an example of our method. Since this methodology will lay the foundations for future studies on the functional form of $c_s^2$ of neutron stars, we also make a Mathematica code that can reproduce these EOS available at \href{https://github.com/jnoronhahostler}{https://github.com/jnoronhahostler}. 

One consequence of the large jump in $c_s^2$ that is discussed in this paper is that the maximum mass of a neutron star for our family of EOS occurs at relatively low $n_B/n_{sat}$.  In the plots below a black dot is shown at the points in the functional form of $c_s^2$ where the central density of the neutron star is reached.  Beyond this density neutron stars are no longer stable. Most of our EOS have a maximum central density of $n_B/n_{sat}=2-4$ with a handful of extreme EOS that can have a central density as low as $n_B/n_{sat}=1.5$ or a high one at $n_B/n_{sat}=5$.  These extreme cases are only reached with the speed of sound contains a dramatic jump up to around the causal limit either at low or high densities. 

%
\begin{figure*}
\centering
\begin{tabular}{c c}
\includegraphics[width=0.5\linewidth]{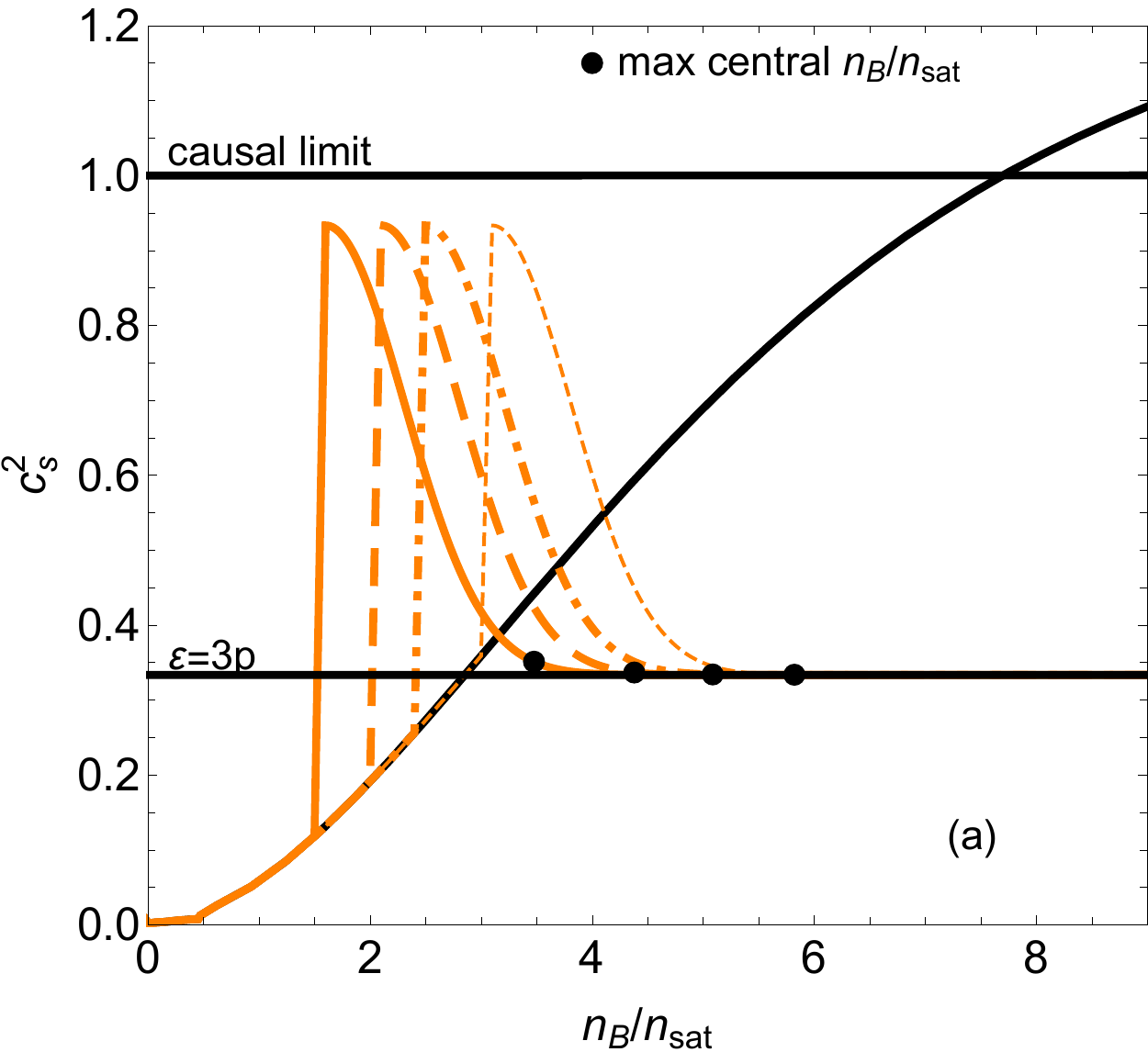} & \includegraphics[width=0.5\linewidth]{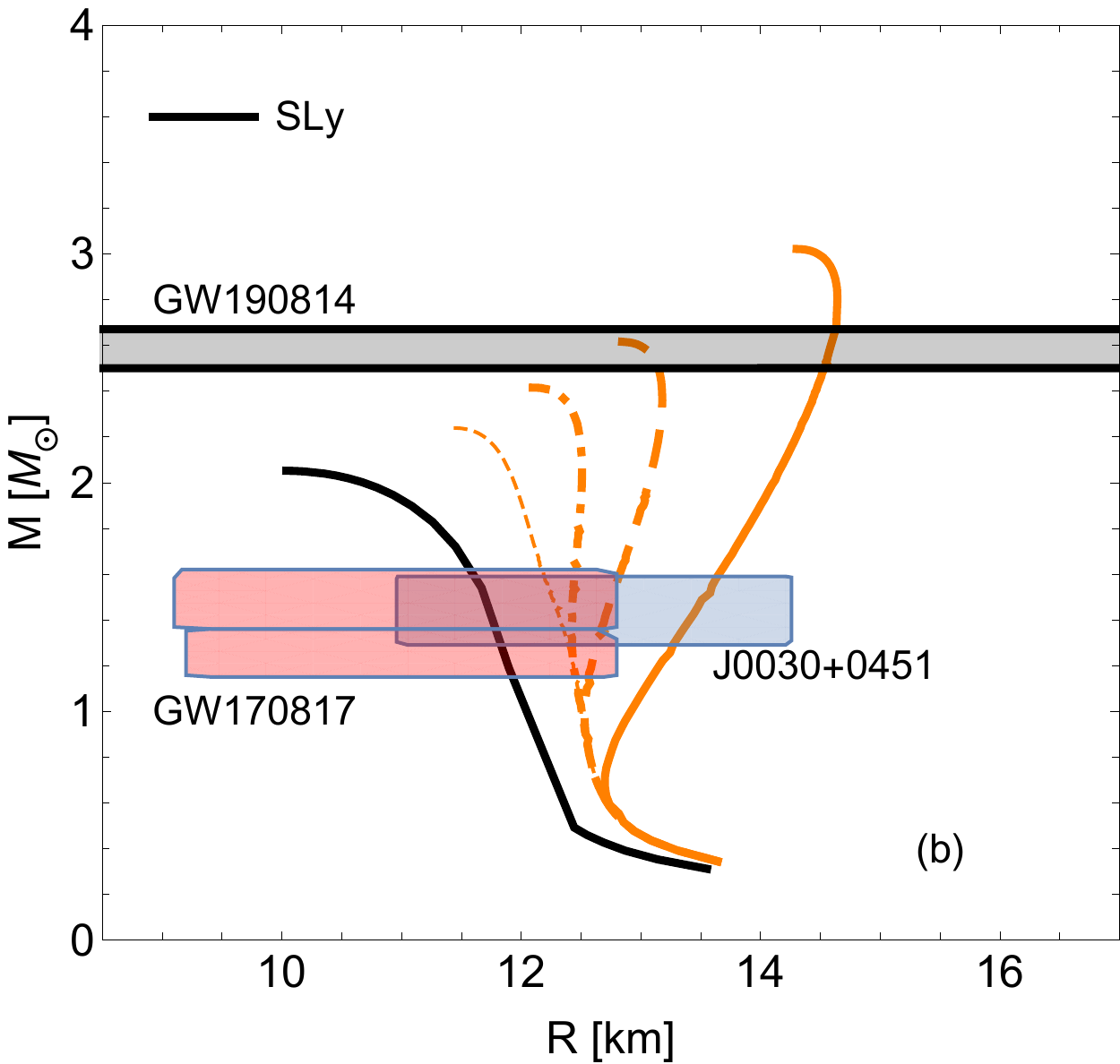}
\end{tabular}
\caption{(Color online) Speed of sound (a) and mass-radius curve (b) for a sub-family of EOSs with peaks in $c_s^2$ peaks of the same width at different locations. The parameters of this subfamily were chosen to  guarantee that the EOS remain causal and produce some  neutron stars with $M_{max}\geq 2.5 M_\odot$. Observe that peaks at low baryon densities have the effect of increasing the maximum mass and the average radius.}
\label{fig:peak}
\end{figure*}
As an example of a sub-family of EOSs we study one which contains a peak in the speed of sound, similar to what was studied in~\cite{Tews:2018kmu}, which can also be well motivated by quarkyonic matter~\cite{McLerran:2018hbz}.  The functional form of the speed of sound used to create this EOS is\\
\begin{align}\label{eqn:1bump}
c_s^2(n_B)=
\begin{cases}
                                   f_1^{sly}(\frac{n_B}{n_{sat}})  & n_B\leq n_1 \\
                                   f_2(n_B)\equiv (\frac{n_B-n_{min}}{n_{sat}})^2+c_0 & n_1<n_B<n_2 \\
   f_3(n_B)\equiv c_{end}-m \exp\left[-\frac{(n_B/n_{sat}-s)^2}{w^2}\right] & n_B\geq n_2
  \end{cases}
  \end{align}
where $n_{min}$ and $c_0$ are determined by solving the following equations given the transition densities $n_1$ and $n_2$:
\begin{eqnarray}
n_{min} &=&\frac{ f_2(n_2)-f_1(n_1)+\left[(n_1/ n_{sat})^2-(n_2/n_{sat})^2\right]}{2(n_1/n_{sat}-n_2/n_{sat})} n_{sat}\nonumber\\
c_0 &=&  f_2(n_2)-(n_2-n_{min})/n_{sat}
\end{eqnarray}
and $c_{end}$ is the final $c_s^2$ at large densities, $w$ is the width of the peak, $m$ provides a magnitude for the effect, and $s$ shifts the position of the peak. This sub-family of EOSs is then parameterized by $\left\{n_1,n_2,c_{end},m,s,w\right\}$.

Figure\ \ref{fig:peak} shows the effect that a single peak in the speed of sound at different locations has on the EOS and the resulting mass-radius relation. An increase in the transition density, $n_1$, at which the peak occurs has the effect of decreasing the maximum mass, while also decreasing the average radius. As the peak is shifted to higher $n_B/n_{sat}$ it is clear that the central density that leads to the maximum mass also increases.  However, for this specific peak eventually placing it at too high of $n_B/n_{sat}\sim 2.5$ leads to a maximum mass that is too small.  One can see that a peak that leads to a sufficiently high maximum mass has a maximum central density that is relatively low.  This implies that degrees of freedom that would be relevant at higher baryon densities would not be possible to probe using the mass-radius relationship of a neutron star. Finally, we point out the the maximum central density is directly related to the radius of the star i.e. a low maximum central density implies a ``fluffier" star with a large radius and a higher maximum central density implies a compact start with a small radius.  We find that EOS that can produce $M>2.5 M_\odot$ generally are less dense and, therefore, produce large radii.

\end{document}